\documentclass[aps,prb,preprint,superscriptaddress,showpacs]{revtex4-1}
\usepackage{graphicx}
\usepackage{amssymb}
\usepackage{amsmath}
\usepackage{amsfonts}
\usepackage{latexsym}
\usepackage[usenames,dvipsnames]{color}

\begin{document}


\title{Electron-phonon coupling and two-band superconductivity \\
       of Al- and C-doped MgB$_2$}

\author{O. De la Pe\~na-Seaman}
\affiliation{Departamento de F{\'\i}sica Aplicada, Centro de Investigaci\'on 
             y de Estudios Avanzados del IPN\\
        Apartado Postal 73 Cordemex 97310 M\'erida, Yucat\'an, M\'exico}
\affiliation{Karlsruher Institut f\"{u}r Technologie (KIT), Institut f\"{u}r 
             Festk\"{o}rperphysik \\
             P.O. Box 3640, D-76021 Karlsruhe, Germany}
                     
\author{R. de Coss}
\affiliation{Departamento de F{\'\i}sica Aplicada, Centro de Investigaci\'on 
             y de Estudios Avanzados del IPN\\
        Apartado Postal 73 Cordemex 97310 M\'erida, Yucat\'an, M\'exico}

\author{R. Heid}
\affiliation{Karlsruher Institut f\"{u}r Technologie (KIT), Institut f\"{u}r 
             Festk\"{o}rperphysik \\
             P.O. Box 3640, D-76021 Karlsruhe, Germany}
             
\author{K.-P. Bohnen}
\affiliation{Karlsruher Institut f\"{u}r Technologie (KIT), Institut f\"{u}r 
             Festk\"{o}rperphysik \\
             P.O. Box 3640, D-76021 Karlsruhe, Germany}
                     
\date{December 10, 2010}

\begin{abstract} 
We have studied the electron-phonon and superconducting properties of the 
Mg$_{1-x}$Al$_x$B$_2$ and MgB$_{2(1-y)}$C$_{2y}$ alloys within the framework 
of density functional theory using the self-consistent virtual-crystal 
approximation. For both alloys, the Eliashberg spectral functions and the 
electron-phonon coupling constants have been calculated in the two-band model 
for several concentrations up to $x$(Al)$=0.55$ and $y$(C)$=0.175$. 
We solved numerically the two-band Eliashberg gap equations without
considering interband scattering. 
Using a single parameter for the Coulomb pseudopotential, which was
determined for the undoped compound, we were able to reproduce the
experimental doping dependence of $\Delta_{\sigma}$, $\Delta_{\pi}$,
and $T_c$ for both alloys on a quantitative level. 
In particular, the observed differences in the doping range of 
superconductivity between Al and C doping indicate a pronounced influence of 
the doping site, which can be explained naturally in the present approach 
without the need to invoke interband scattering, suggesting that this factor 
plays only a minor role.
\end{abstract}


\maketitle

\section{INTRODUCTION} 

The discovery of superconductivity in 2001 in the intermetallic compound 
MgB$_2$, with a critical temperature $T_c \approx 39$ K \cite{naga}, has 
motivated a lot of theoretical and experimental studies in order to understand 
the origin and characteristics of the relatively high $T_c$ in this material. 
It is now generally accepted that MgB$_2$ is a phonon-mediated
superconductor, and that the high transition temperature arises due to a
combination of several peculiar features in its electronic structure
and electron-phonon coupling, which conspire to produce a superconducting
state with multiple gaps
\cite{an,kortus,baron2,flor2,kortus2,szab,iava,schmd,gonne4}. Its electronic 
band structure in the vicinity of the Fermi energy consists of two bonding 
$\sigma$ bands corresponding to in-plane $s-p_x-p_y$ ($sp^2$) hybridization in 
the boron layer and two $\pi$ bands (bonding and anti-bonding) formed by 
hybridized boron $p_z$ orbitals. A substantial part of the electron-phonon 
coupling has its origin in the interaction of states at the $\sigma$-band Fermi 
surfaces with one specific phonon mode, the B-B bond stretching mode with 
$E_{2g}$ symmetry at the $\Gamma$ point 
\cite{kong,bohn1,yild,kunc,baron2,kortus2}. In addition, MgB$_2$ possesses two 
distinct superconducting gaps associated with the $\sigma$ and $\pi$ Fermi 
surfaces. This superconducting state can be described within a multiband 
version of the Eliashberg theory where the pairing interaction is split into 
intra- and interband contributions \cite{liu,kong,golu,geerk2}.

In the search for related compounds with similar outstanding superconducting
properties, only a few variants of MgB$_2$ have been found.
Among them, two alloy systems could be  successfully synthesized based on the 
partial substitution of Mg by Al \cite{slusky,bian,posto,yang} and B by C 
\cite{take,ribe,avde,kaza,lee}, respectively. Both substitutions provide 
electron doping to the alloy and lead to a reduction of $T_c$. For 
Mg$_{1-x}$Al$_x$B$_2$ and MgB$_{2(1-y)}$C$_{2y}$, loss of superconductivity is 
found for $x>0.5$ \cite{slusky,bian,posto,yang} and $y>0.15$ 
\cite{take,ribe,avde,kaza,lee}, respectively. A correlation between the $T_c$ 
reduction and the filling of the hole-type $\sigma$-bands as a function of 
Al-doped content was found by {\em first principles} calculations within the 
virtual-crystal approximation \cite{omar,omar4}.
In parallel with the reduction of $T_c$ also a decrease of the superconducting 
gaps $\Delta_{\sigma}$ and $\Delta_{\pi}$ has been observed. For single 
crystals as well as polycrystals Al doping was found to decrease both $\sigma$ 
and $\pi$ gaps monotonically, which, however, stay distinguishable even for 
$T_c$ as low as 10 K $(x \approx 0.32)$ 
\cite{gonne,karpi,gonne3,klein,szab2,gonne2,schne}. These observations indicate 
that the interband scattering $\Gamma_{\sigma\pi}$ remains small even at high 
doping levels, and is insufficient to produce a merging of the $\sigma$ and 
$\pi$ gaps.
On contrast, for the C-doped system, contradictory experimental results have 
been reported with respect to the question if the superconducting gaps merge as 
a function of doping \cite{gonne3,schmd2,hola,tsud,szab2}. In all experiments a 
decrease of both gaps with C doping was observed. There are point-contact 
tunneling \cite{schmd2,hola}, point-contact spectroscopy \cite{szab2} and 
photoemission spectroscopy \cite{tsud} measurements that show a clear 
difference between the $\sigma$ and $\pi$ gaps at all doping levels. However, 
there exists also point-contact spectroscopy measurements \cite{gonne3} that 
suggest a merging of the gaps at $T_c \approx 17$ K ($y \approx 0.13$). This 
was then interpreted as a doping-induced increase of the interband scattering 
$\Gamma_{\sigma\pi}$, which tends to reduce gap anisotropies.

From the theoretical point of view, many investigations have been performed to 
study the doping dependence of the structural \cite{bara,omar}, electronic 
\cite{suzu,omar,ppsingh2,kasi,cool,choi3}, vibrational \cite{renk,profe,moud,
choi3} and superconducting properties \cite{ppsingh1,ppsingh2,buss,profe,umma2,
umma,kortus3,ppsingh3,choi3} of the Mg$_{1-x}$Al$_x$B$_2$ and 
MgB$_{2(1-y)}$C$_{2y}$ systems using different approximations for the 
simulation of the alloys, like the supercell approach \cite{bara,suzu,buss,
renk,moud}, the rigid band approximation (RBA) \cite{choi3}, the virtual-crystal 
approximation (VCA) \cite{omar,profe,kortus,cool}, the coherent potential 
approximation (CPA) \cite{ppsingh1,ppsingh2,kasi}, and the 
Korringa-Kohn-Rostoker coherent potential approximation (KKR-CPA) 
\cite{ppsingh3}. However, in particular for the supercell and CPA approaches, 
the studies have been limited to a few Al or C concentrations only, because 
these calculations are computationally very demanding, especially if one is 
interested in very low (close to Mg or B) or high (close to Al) concentrations. 
Additionally, in these approaches the symmetry of the original system is lost, 
which complicates the interpretation and understanding of experimental results 
as a function of doping. 

In this paper, we present a study of the electron-phonon and superconducting 
properties of Mg$_{1-x}$Al$_x$B$_2$ and MgB$_{2(1-y)}$C$_{2y}$ within the 
framework of density functional theory \cite{kohn} using the self-consistent 
virtual-crystal approximation (VCA) \cite{mehl,omar,omar2,omar3}. We calculate 
the electron-phonon (e-ph) properties such as the Eliashberg function, 
$\alpha^{2}_{ij}F(\omega)$, and the e-ph coupling constant, $\lambda_{ij}$, 
within the two-band model as a function of doping. By solving the two-band 
Eliashberg gap equations on the imaginary axis we obtain the superconducting 
gaps, $\Delta_{\sigma}$ and $\Delta_{\pi}$, and the value of $T_c$ as a 
function of $x$ or $y$ for the Mg$_{1-x}$Al$_x$B$_2$ and MgB$_{2(1-y)}$C$_{2y}$ 
alloys, respectively. The evolution of these quantities is analyzed and 
discussed in connection with changes in the electronic and vibrational 
properties.

\section{COMPUTATIONAL DETAILS} 

The calculations were performed with the mixed-basis pseudopotential method 
(MBPP) \cite{louie,meyer}. For Mg/Al and B/C norm-conserving pseudopotentials 
were constructed according to the Vanderbilt description \cite{vander}. Details 
of pseudopotentials, basis functions and calculational aspects for ground-state 
and phonon properties  can be found in a previous publication \cite{omar4}. The 
Mg$_{1-x}$Al$_x$B$_2$ and MgB$_{2(1-y)}$C$_{2y}$ alloys were modeled in the 
self-consistent virtual-crystal approximation (VCA) \cite{papas,mehl,omar,
ambrox,thon,omar2,omar3,omar4}. The VCA is implemented within the MBPP method 
\cite{louie,meyer} by generating new pseudopotentials with a fractional nuclear 
charge at the Mg or B site for each $x$ and $y$, respectively (Al: $Z$=12+$x$ 
and C: $Z$=5+$y$), and by adjusting the valence charge accordingly 
\cite{omar4}. From our previous results for the electronic and vibrational 
properties \cite{omar4} the screened electron-phonon matrix elements were 
calculated via density functional perturbation theory 
\cite{kohn,baro,gian,heid}, which are the key elements of the Eliashberg theory 
\cite{eliash,carb,sc01,savra2}. The calculations employ the PBE version of the 
generalized gradient approximation (GGA) \cite{ozo,perdew3,kokko}, and are 
performed at the optimal lattice parameters for each doping level \cite{omar4}.
Eliashberg functions for all band combinations where obtained by standard 
Fourier interpolation of quantities calculated with a dense 
$36\times36\times36$ $k$-point mesh and a $6\times6\times6$ $q$-point mesh. The 
original four-band Eliashberg functions are projected onto an effective 
two-band model by averaging over the two $\sigma$ and the two $\pi$ bands, 
respectively. The partial and total Eliashberg functions are given by the 
following expressions,

\begin{equation}
\alpha_{ij}^2F(\omega)=\frac{1}{N_i}\sum_{{\bf q}\nu}\delta(\omega-\omega_{\bf{q}\nu})
\sum_{{\bf k},{\bf k}_n}|g_{{\bf k},i,{\bf k}_n,j}^{{\bf q}\nu}|^2
\delta(\epsilon_{{\bf k},i}-\epsilon_F)
\delta(\epsilon_{{\bf k}_n,j}-\epsilon_F),
\end{equation}

\begin{equation}
\alpha^2F(\omega)=\frac{1}{N_{tot}}\sum_{ij}N_i\alpha_{ij}^2F(\omega),
\end{equation}

\noindent
where $i$ and $j$ are the band indices $\sigma$ or $\pi$, $N_i$ ($N_{tot}$) is 
the partial (total) electronic density of states at the Fermi level (per atom 
and spin), and $g_{{\bf k},i,{\bf k}_n,j}^{{\bf q}\nu}$ is the e-ph matrix 
element for scattering of an electron from a Bloch state with momentum 
${\bf k}$ to another Bloch state ${\bf k}_n = {\bf k} + {\bf q}$ by a phonon 
${\bf q}\nu$ ($\nu$ indicates the branch index and $\omega_{\bf{q}\nu}$ is the 
phonon frequency).

In a similar way, the partial and total e-ph coupling parameters ($\lambda$) 
are expressed as follows,

\begin{equation}
\lambda_{ij}=2\int \frac{d\omega}{\omega} \alpha_{ij}^2F(\omega),
\end{equation}

\begin{equation}
\lambda_{tot}=\frac{1}{N_{tot}}\sum_{ij}N_i\lambda_{ij},
\end{equation}

Within the two-band model, there are three independent contributions to 
$\alpha^{2}_{ij}F(\omega)$: two intraband ($\pi\pi$ and $\sigma\sigma$) and one 
interband $\alpha^{2}_{\pi\sigma}F(\omega)=\frac{N_{\sigma}}{N_\pi}
\alpha^{2}_{\sigma\pi}F(\omega)$. With the knowledge of 
$\alpha^{2}_{ij}F(\omega)$, the two-band Eliashberg gap equations 
\cite{kres,eliash,carb,sc01} on the imaginary axis were numerically solved in 
order to obtain the gap values  and $T_c$ for each given Al or C concentration, 
respectively. This procedure has been previously used in similar studies of 
undoped MgB$_2$ \cite{golu,umma2}. The solution involves four non-linear 
coupled equations for the Matsubara gaps $\Delta_i(i\omega_n)$ and the 
renormalization functions $Z_i(i\omega_n)$,

\begin{equation}
\Delta_i(i\omega_n)Z_i(i\omega_n)=\pi T\sum_{m,j}\left[
\Lambda_{ij}(i\omega_m-i\omega_n)-\mu^{*}_{ij}(\omega_c)\theta(\omega_c-|\omega_m|) 
\right]N^{j}_{\Delta 1}(i\omega_m),
\end{equation}

\begin{equation}
Z_i(i\omega_n)=1+\frac{\pi T}{\omega_n}\sum_{m,j}\Lambda_{ij}(i\omega_m-i\omega_n)
N^{j}_{\Delta 0}(i\omega_m),
\end{equation}

\noindent
where $\theta$ is the Heaviside function, and $\mu^{*}_{ij}$ is the Coulomb 
pseudopotential, $\omega_c$ a cutoff frequency (chosen as 
$\omega_c\approx 10\omega^{max}_{ph}$) and $\omega_n=\pi T(2n-1)$, with 
$n=0,\pm 1,\pm 2,\ldots$, is the discrete set of Matsubara frequencies. The 
pairing interaction is contained in the kernel 

\begin{equation}
\Lambda_{ij}(i\omega_m-i\omega_n)=2\int^{\infty}_{0}
\frac{\omega\alpha^{2}_{ij}F(\omega)d\omega}{\omega^2+(\omega_n-\omega_m)^2},
\end{equation}
and we defined the following quantities:
\begin{equation}
N^{j}_{\Delta 1}(i\omega_m)=
\frac{\Delta_j(i\omega_m)}{\sqrt{\omega^{2}_{m}+\Delta^{2}_{j}(i\omega_m)}},
\end{equation}

\begin{equation}
N^{j}_{\Delta 0}(i\omega_m)=
\frac{\omega_m}{\sqrt{\omega^{2}_{m}+\Delta^{2}_{j}(i\omega_m)}}.
\end{equation}

In order to keep the number of adjustable parameter to a minimum, we 
approximated the Coulomb pseudopotential matrix $\mu^*_{ij}$, which is a 
two-by-two matrix in the case of the two-band model, by a diagonal form 
proposed earlier \cite{mazin2} as $\mu^*_{ij}=\mu^*_0 \delta_{ij}$. 
\footnote{In some previous works (Refs. \onlinecite{choi,choi4,choi5}) also an 
uniform matrix form for the Coulomb pseudopotential matrix has been adopted. It 
has been demonstrated that this specific form underestimates considerably the 
value of $\Delta_{\pi}$ (Ref. \onlinecite{mazin2}). However, we tested another 
form used in previous publications (Refs. \onlinecite{golu,umma}) where the 
diagonal elements are considered to be different between them additionally 
off-diagonal ones are included, nevertheless they are much smaller than the 
diagonal ones. Results in Section III were practically the same for both forms, 
except for a slight reduction of $\Delta_{\pi}$ by 10\%.}
The gap values were identified with 
$\Delta_i(i\omega_1)$, which corresponds to the point on the imaginary axis 
which is closest to the real axis. Test calculations solving the Eliashberg 
equations on the real axis indicated that this approximation is accurate on the 
level of 1\% or better.

\section{RESULTS AND DISCUSSION}

Based on our previous results for the electronic and vibrational properties 
\cite{omar4}, we have calculated the electron-phonon coupling quantities of the 
two-band model ($\alpha^{2}_{ij}F(\omega)$ and $\lambda_{ij}$ with 
$i,j=\sigma,\pi$) for the ranges $x \leqslant 0.55$ and $y \leqslant 0.175$ in 
the Al- and C-doped systems, respectively.


\begin{figure}[htbp]
\centering
\vspace*{0.0 true cm}
\includegraphics*[scale=0.84,trim=2mm 15.0cm 0mm 0mm,clip]{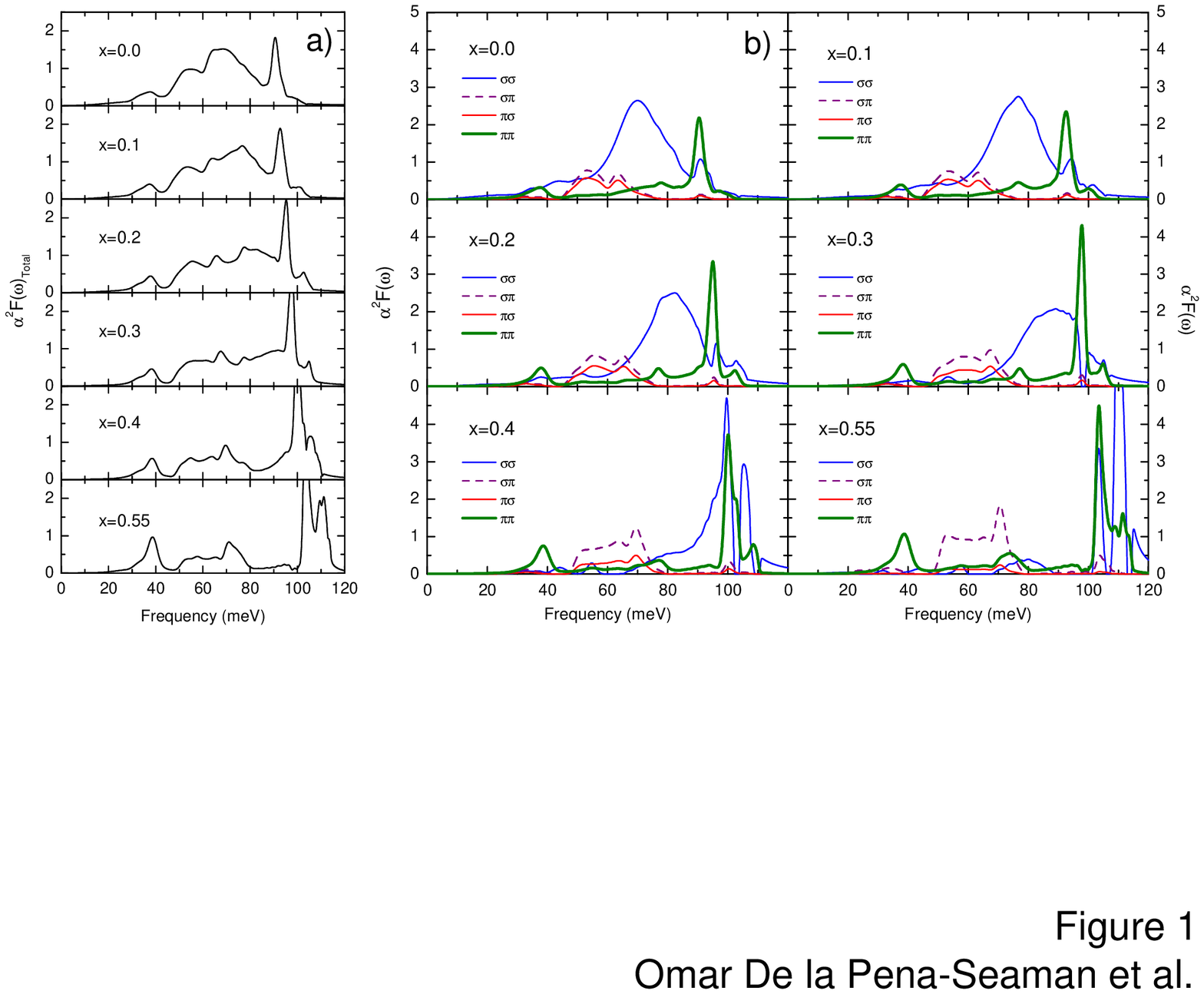}
\vspace*{-1.5 true cm}
\caption{\label{fig:fig01}(Color online) Evolution of the (a) total Eliashberg 
         function and (b) components $\alpha^{2}_{ij}F(\omega)$ 
         ($i,j=\sigma,\pi$) for the Mg$_{1-x}$Al$_x$B$_2$ alloy.}
\end{figure}

In Fig. 1 we show the evolution of the Eliashberg functions, the total spectra 
and the four components $\alpha^{2}_{ij}F(\omega)$ ($i,j=\sigma,\pi$) of 
Mg$_{1-x}$Al$_x$B$_2$ for six Al concentrations in the superconducting regime 
($x=$0.0, 0.1, 0.2, 0.3, 0.4, and 0.55). 
We observe that for MgB$_2$ ($x=0$) the largest contribution to the total 
spectral function comes from the $\sigma\sigma$ component, where the main peak 
centered at approximately 70 meV corresponds to frequency of the 
$E_{2g}$-phonon mode. The $\pi\pi$ spectrum has its main contribution from the 
high-frequency phonon region related to the $B_{1g}$-phonon mode, while the 
interband contribution, $\sigma\pi$($\pi\sigma$), is concentrated in the region 
between 50 and 70 meV.
We note that, although the $\sigma\sigma$ part represents the main contribution 
to the total spectra, the other components can not be neglected in a proper 
quantitative description of the e-ph coupling and of the superconducting 
properties.


\begin{figure}[htbp]
\centering
\vspace*{0.0 true cm}
\includegraphics*[scale=0.84,trim=2mm 15.0cm 0mm 0mm,clip]{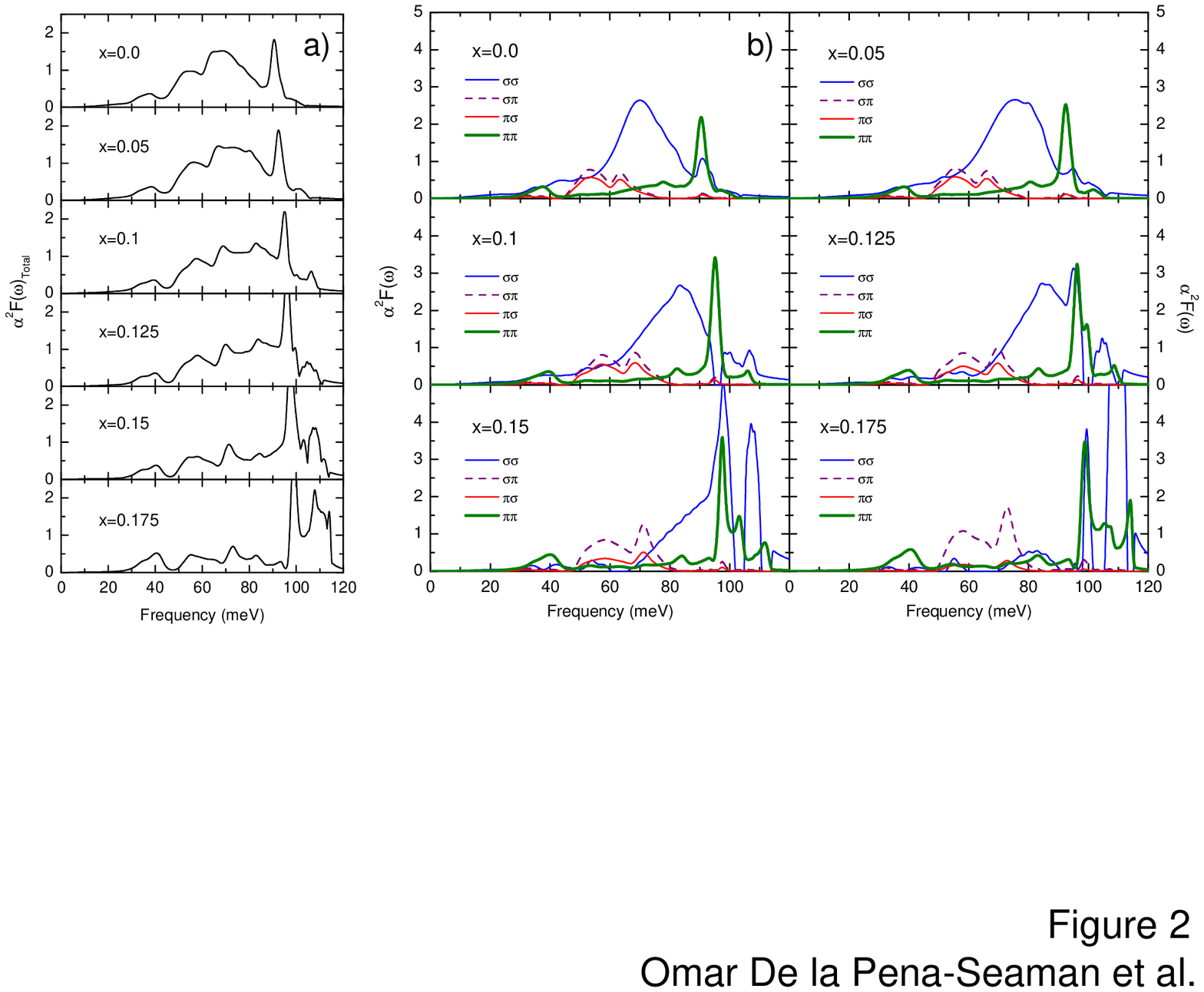}
\vspace*{-1.5 true cm}
\caption{\label{fig:fig02} (Color online) Evolution of the (a) total Eliashberg 
         function and (b) components $\alpha^{2}_{ij}F(\omega)$ 
         ($i,j=\sigma,\pi$) for the MgB$_{2(1-y)}$C$_{2y}$ alloy.}         
\end{figure}

From the evolution of spectral functions for Mg$_{1-x}$Al$_x$B$_2$ (Fig. 1) we 
observe that almost all components are reduced by Al-doping, but the largest 
changes are exhibited by $\alpha^{2}_{\sigma\sigma}F(\omega)$. Its main peak  
shifts to higher frequencies and its area is reduced at the same time until it 
practically vanishes for $x=0.55$. This doping level is close to the region 
where the loss of superconductivity has been observed experimentally 
($x \gtrsim 0.5$) \cite{slusky,bian,posto,yang}. The reduction of 
$\alpha^{2}_{\sigma\sigma}F(\omega)$ indicates the loss of intraband e-ph 
coupling between the $\sigma$ states and the bond-stretching phonon modes and 
has its origin in the continuous filling of the $\sigma$ bands, which is 
completed at $x_c =0.57$ \cite{omar4}. The shift to higher frequencies is 
due to the hardening of the $E_{2g}$-phonon mode as $x$ increases, 
a phenomenon discussed previously \cite{omar4}. Similar to $\sigma\sigma$, the 
$\pi\sigma$ interband contribution is also reduced as a function of $x$ and 
almost disappears at $x=0.55$. On contrast, the $\sigma\pi$ contribution shows 
an slight increase around 50 and 70 meV, and the $\pi\pi$ contribution at 
higher frequencies strengthens slightly with doping, but the position of its 
main peak is almost unaffected. In recent electron tunneling spectroscopy 
measurements on Al-doped thin films \cite{schne}, this general behavior of the 
Eliashberg function indeed has been observed, supporting our results.


\begin{figure}[htbp]
\centering
\vspace*{-0.05 true cm}
\includegraphics*[scale=0.80,trim=2mm 13.4cm 0mm 0mm,clip]{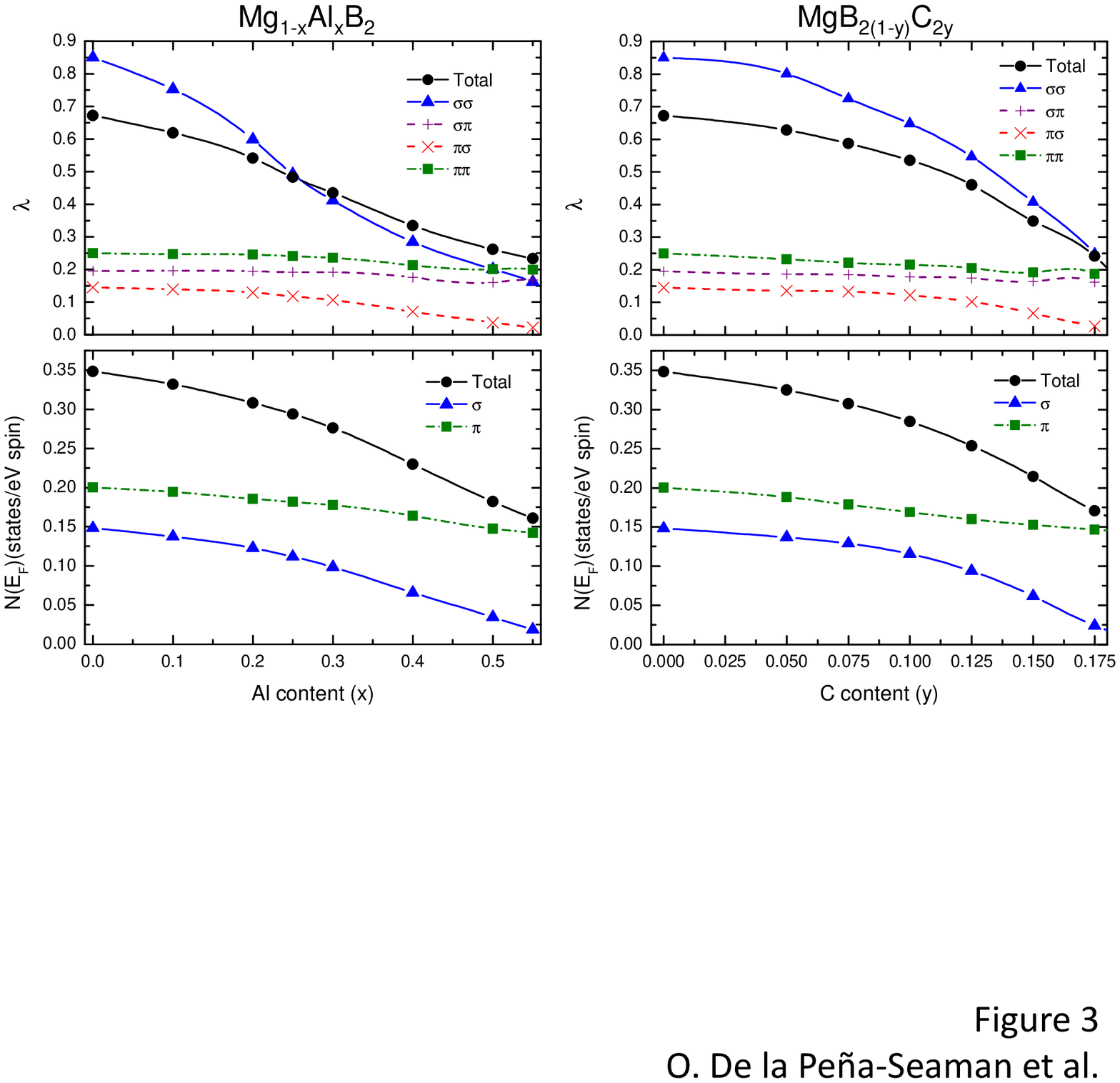}
\vspace*{-1.4 true cm}
\caption{\label{fig:fig03} (Color online) Evolution of $\lambda_{ij}$ and 
         $N_i(E_F)$ as a function of $x$ and $y$ for Mg$_{1-x}$Al$_x$B$_2$ and 
         MgB$_{2(1-y)}$C$_{2y}$, respectively.}
\end{figure}


In Fig. 2 we present the results for the Eliashberg functions of 
MgB$_{2(1-y)}$C$_{2y}$ for six C concentrations in the superconducting region 
of the alloy ($y=$0.0, 0.05, 0.1, 0.125, 0.15, and 0.175). The different 
components of $\alpha^{2}F(\omega)$ exhibit the same trends with increasing C 
concentration as those found for the Al-doping. However the shape of the 
spectras are different and the changes take place at lower concentrations. 
When comparing the two alloys, one should take into account that the number of 
doping-induced electrons per unit cell is given by $x$ and $2y$, respectively. 
Even with this factor of two, the vanishing of the $\sigma\sigma$ and 
$\sigma\pi$ contributions at $2y \approx 0.35$ occurs at a much smaller doping 
level than for Al-doping ($x=0.55$). The dramatic reduction of 
$\alpha^{2}_{\sigma\sigma}F(\omega)$ at $y \approx 0.175$ correlates also with 
the complete filling of the $\sigma$-bands on MgB$_{2(1-y)}$C$_{2y}$ 
\cite{omar4}. 

In Fig. 3 calculated total and partial contributions for $\lambda$ as well as 
for $N(E_F)$ are shown. For MgB$_2$ the calculated values are 
$\lambda_{\sigma\sigma}=0.850$, $\lambda_{\sigma\pi}=0.196$, 
$\lambda_{\pi\sigma}=0.145$, $\lambda_{\pi\pi}=0.250$, and 
$\lambda_{tot}=0.672$. It is worth mentioning that this $\lambda_{tot}$ value 
is very close to the experimental one by Geerk {\it et al.} \cite{geerk2} 
($\lambda_{eff}=0.650$). $N(E_F)$ partial contributions for MgB$_2$ are 
$N_{\sigma}=0.148$ states eV$^{-1}$/spin and $N_{\pi}=0.200$ states 
eV$^{-1}$/spin, which are very similar to those calculated earlier by Liu 
{\it et al.} \cite{liu} and Golubov {\it et al.} \cite{golu}. As seen from Fig. 
3, the main contribution to the e-ph coupling ($\lambda$) in undoped MgB$_2$ 
comes from the $\sigma\sigma$ component. 
Among the different contributions of the e-ph coupling, $\lambda_{\sigma\sigma}$ 
shows the largest changes on doping with a reduction of $\approx 75\%$ 
(comparing the boundary concentrations). The other components also decrease 
with doping, albeit at different scales, ranging from the nearly constant 
behavior of $\lambda_{\sigma\pi}$ and $\lambda_{\pi\pi}$ to an almost complete 
vanishing of $\lambda_{\pi\sigma}$. As a consequence, $\lambda_{tot}$ 
monotonically decreases with doping.


\begin{figure}[htbp]
\centering
\vspace*{-0.05 true cm}
\includegraphics*[scale=0.535,trim=2mm 5.4cm 0mm 0mm,clip]{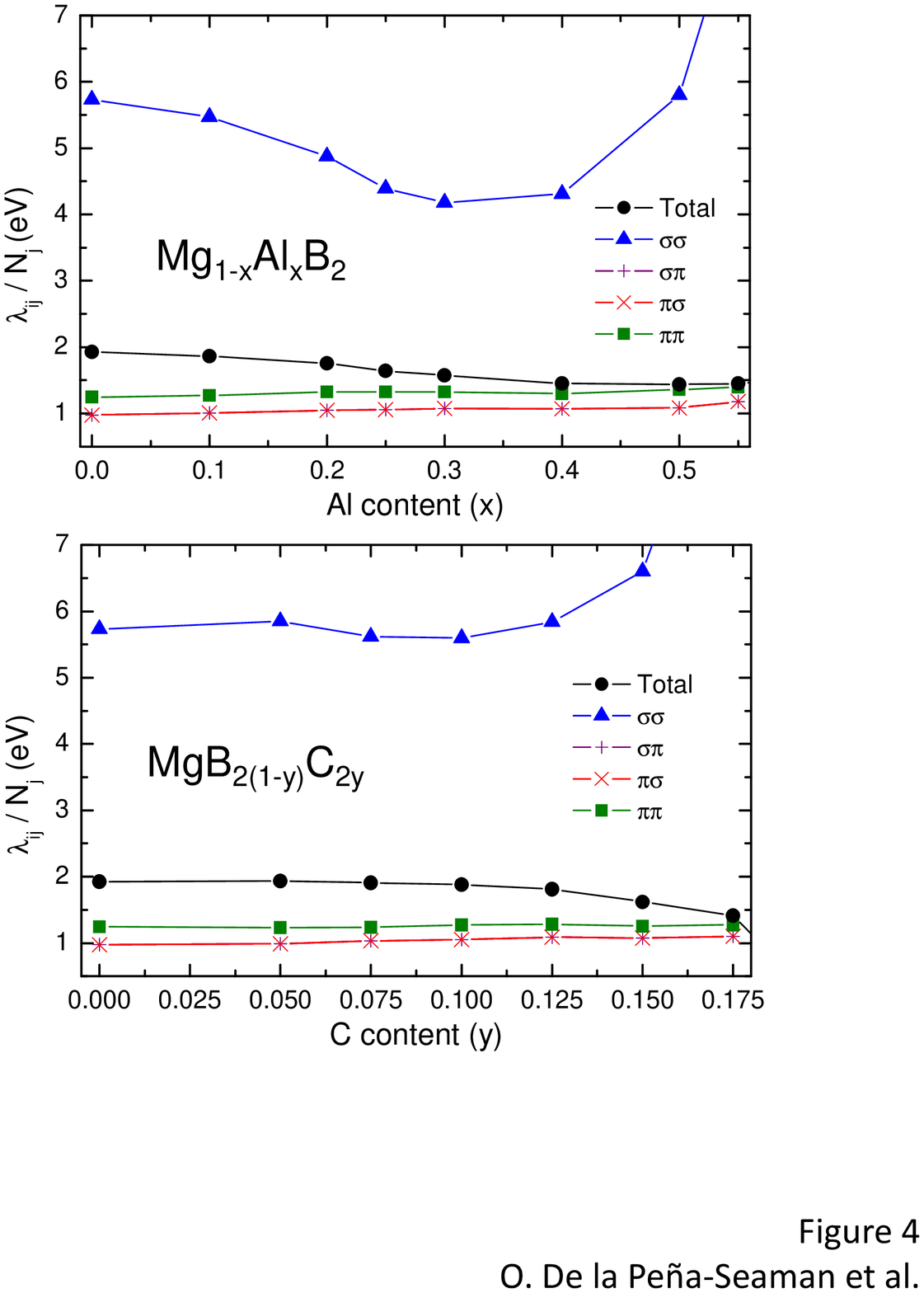}
\vspace*{-1.4 true cm}
\caption{\label{fig:fig04} (Color online) Evolution of $\lambda_{ij} / N_j$ and 
         $\lambda_{tot}/N_{tot}$ as a function of $x$ and $y$ for 
         Mg$_{1-x}$Al$_x$B$_2$ and MgB$_{2(1-y)}$C$_{2y}$, respectively. Note 
         that $\lambda_{\pi\sigma} / N_\sigma\equiv \lambda_{\sigma\pi}/N_\pi$.}
\end{figure}


Doping-induced changes in the coupling constants can arise from changes in the 
partial density of states or from changes in the e-ph matrix elements. In order 
to distinguish between these two possibilities, we plotted in Fig. 4 the ratios 
$\lambda_{ij}/N_{j}$ and $\lambda_{tot}/N_{tot}$.
Indeed, a relationship $\alpha_{ij}^2F(\omega) \sim N_j$ and $\lambda_{ij}
\sim N_j$ can easily be derived from Eq. 1 under the assumption of 
momentum-independent e-ph matrix elements. The ratios $\lambda_{ij} / N_j$ for 
the interband ($\lambda_{\sigma\pi}$, $\lambda_{\pi\sigma}$) as well as for the 
intraband $\lambda_{\pi\pi}$ couplings remain practically constant as a 
function of doping for both alloys, indicating that the corresponding e-ph 
matrix elements are approximately independent of doping.
However, for the ($\sigma\sigma$) intraband coupling, the ratio exhibits a 
stronger variation with doping, in particular for Al doping, which signals a 
clear doping dependence of the e-ph matrix elements.
In this case, a simple scaling with the partial density of states would be 
inappropriate to describe the doping dependence of $\lambda_{\sigma\sigma}$.


\begin{figure}[htbp]
\centering
\vspace*{0.0 true cm}
\includegraphics*[scale=0.60,trim=0mm 11.5cm 0mm 5mm,clip]{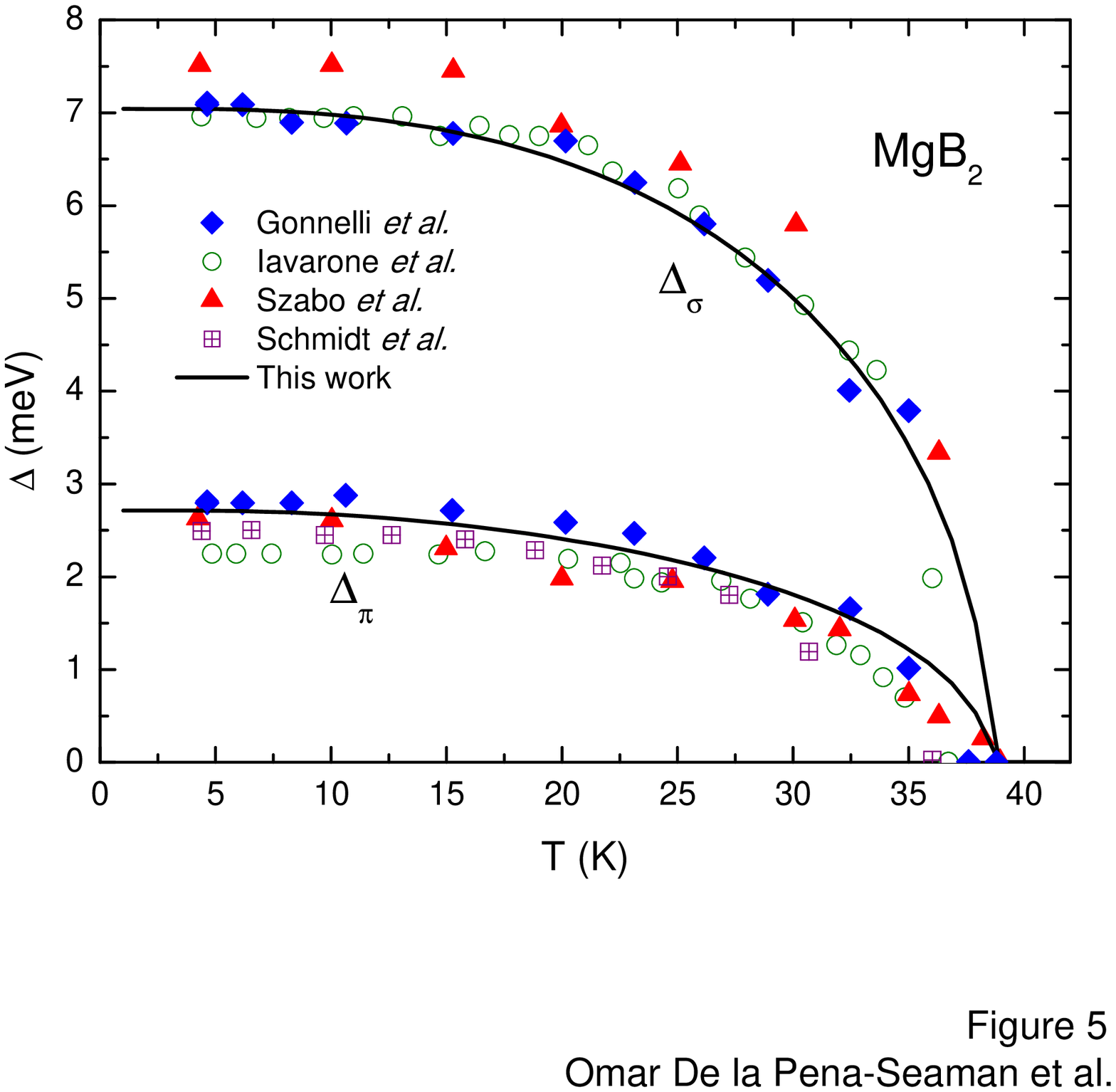}
\caption{\label{fig:fig05} (Color online) Temperature dependence of the 
         superconducting gaps $\Delta_{\sigma}$ and $\Delta_{\pi}$ (solid lines) 
         for undoped MgB$_2$ as obtained from the two-band Eliashberg gap 
         equations. Symbols represent experimental data 
         ({\color{red} $\blacktriangle$})\cite{szab}, 
         ({\color{OliveGreen}$\circ$})\cite{iava}, 
         ({\color{Purple}$\boxplus$})\cite{schmd}, and
         ({\color{blue}$\blacklozenge$})\cite{gonne4,gonne2}. The calculated 
         gap values at $T \rightarrow 0$ K are $\Delta_{\sigma}=7.04$ meV and 
         $\Delta_{\pi}=2.71$ meV.}
\end{figure}


To solve the Eliashberg gap equations, we determined the single remaining
parameter, the Coulomb pseudopotential $\mu^*_0$, by the requirement that for 
undoped MgB$_2$ the experimental transition temperature of $T_c=38.82$K 
\cite{gonne2,gonne3} is reproduced. For a cutoff frequency 
$\omega_c=10\omega^{max}_{ph}$ we found $\mu^{*}_0=0.107$.
The resulting temperature dependence of the superconducting gaps is shown in 
Fig. 5 and compared with available experimental data 
\cite{szab,iava,schmd,gonne2}. The gap values for $T \rightarrow 0$ K are 
$\Delta_{\sigma}=7.04$ meV and $\Delta_{\pi}=2.71$ meV, respectively, in good 
agreement with experimental results \cite{iava,schmd,gonne2}.


\begin{figure}[htbp]
\centering
\vspace*{-0.05 true cm}
\includegraphics*[scale=0.82,trim=2mm 13.3cm 0mm 0mm,clip]{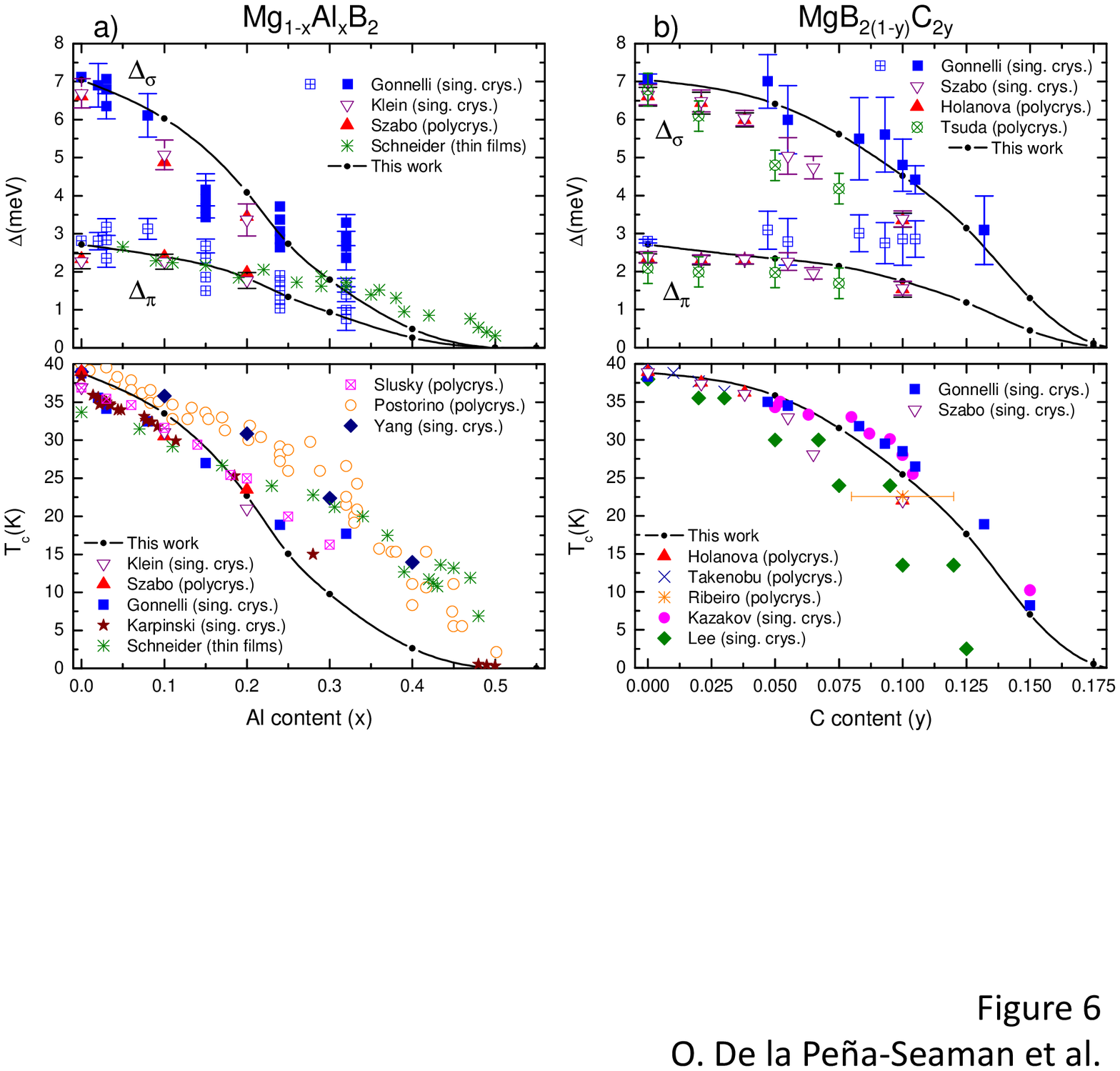}
\vspace*{-1.3 true cm}
\caption{\label{fig:fig06} (Color online) Superconducting gaps $\Delta_{\sigma}$ 
         and $\Delta_{\pi}$ at $T \rightarrow 0$K and critical temperature $T_c$ 
         for (a) Mg$_{1-x}$Al$_x$B$_2$ and (b) MgB$_{2(1-y)}$C$_{2y}$ as a 
         function of $x$ or $y$, respectively. The lines are the present 
         calculations and the symbols represent various experimental 
         measurements; (a): 
         ({\color{magenta} $\boxtimes$})\cite{slusky}, 
         ({\color{Orange} $\circ$})\cite{posto}, 
         ({\color{BlueViolet} $\blacklozenge$})\cite{yang}, 
         ({\color{Sepia} $\bigstar$})\cite{karpi}, 
         ({\color{Purple} $\triangledown$})\cite{klein}, 
         ({\color{red} $\blacktriangle$})\cite{szab2}, 
         ({\color{blue} $\boxplus$,$\blacksquare$})\cite{gonne3,gonne2},
         ({\color{OliveGreen} $\ast$})\cite{schne}; 
         (b): ({\color{RoyalBlue} $\times$})\cite{take}, 
         ({\color{Orange} $\ast$})\cite{ribe,avde}, 
         ({\color{magenta} $\bullet$})\cite{kaza},
         ({\color{OliveGreen} $\blacklozenge$})\cite{lee}, 
         ({\color{Purple} $\triangledown$})\cite{szab2},
         ({\color{blue} $\boxplus$,$\blacksquare$})\cite{gonne3,gonne2}
         ({\color{red} $\blacktriangle$})\cite{hola}, 
         ({\color{OliveGreen} $\otimes$})\cite{tsud}.}
\end{figure}


The same two-band Eliashberg approach was adopted for the alloys
Mg$_{1-x}$Al$_x$B$_2$ and MgB$_{2(1-y)}$C$_{2y}$ keeping $\mu^{*}_0$ at the 
obtained value for undoped MgB$_2$. 
The doping dependence of $\Delta_{\sigma}$, $\Delta_{\pi}$, and $T_c$ for both 
alloys is presented in Fig. 6 and compared with experimental data \cite{slusky,
posto,yang,klein,szab2,karpi,gonne3,gonne2,schne,take,ribe,avde,kaza,lee,hola,
tsud}. The calculations reproduce the experimental trends that both gaps and 
$T_c$ decrease with increasing Al or C doping. 
Beside this common feature the two alloys exhibit also striking differences. 
The first concerns the doping range where superconductivity exists. $T_c$ goes 
to zero close to the critical concentration for which the $\sigma$-band is 
completely filled ($x_c$(Al)$=0.57$, $y_c$(C)$=0.177$) \cite{omar4}. Thus, 
superconductivity vanishes significantly faster on C doping than on Al doping, 
even when taking into account that one should compare doping levels $x=2y$ as 
discussed above. 
The second difference relates to the shape of the $T_c$ versus doping curves. 
For Al doping, $T_c$ initially drops fast and develops a longer tail, whereas 
for C doping $T_c$ is only slowly reduced initially, while it exhibits a 
steeper drop towards the critical concentration where $T_c$ vanishes. A similar 
difference in shape is also observed for the larger gap.
As both alloys are electron-doped systems, these differences indicate the 
importance of the doping site for the superconducting properties. As explained 
in Refs. \onlinecite{omar,omar4}, the origin of this difference can be traced 
back to the distribution of the extra charge introduced by doping. In the 
Al-doped system an important portion of the extra electrons is located in the 
interplanar region, and only a small fraction in the boron planes. In contrast, 
for C doping the extra charge mainly remains in the area between the B atoms 
within the boron plane, exactly in the region of the $\sigma$-bonds. Therefore, 
the extra charge introduced by C doping is more effective in reducing the 
number of holes in the $\sigma$-band and has a stronger influence on the phonon 
frequencies, in particular, on the hardening of the $E_{2g}$ mode. Consequently, 
C doping leads to a faster decrease of the e-ph coupling and of the 
superconducting properties as compared to Al doping.

The various experimental data sets for $T_c$ and the gaps plotted in Fig. 6 
exhibit a clear spread indicating a large dependence of the superconducting 
properties on the sample preparation methods and on the physical conditions of 
the measurement procedure itself.
In addition, an accurate determination of the actual doping concentration in 
these alloys is complicated and far from trivial. Furthermore, there is so far 
no consensus about the behavior of the gaps for larger C doping. While some 
experiments suggest a merging of the $\Delta_{\sigma}$ and $\Delta_{\pi}$ gap 
at $y \approx 0.13$ \cite{gonne3,gonne2}, others find two distinct gaps even 
for the highest doping levels \cite{hola,tsud,szab2}.
Within these experimental uncertainties, our calculations agree quantitatively 
with the data for both alloys. In particular, the different doping regimes are 
obtained in a natural way. We recall that our study involved only a single free 
parameter, $\mu_0^*$, which was fixed for the undoped system and which does not 
directly affect the doping dependence or the gap anisotropy.

In agreement with experimental data, the present calculation predicts for both 
alloys a stronger influence of doping on the $\sigma$ gap, which follows 
approximately the doping dependence of $T_c$. On contrast, the $\pi$ gap 
remains rather stable and only slowly decreases on doping.
This is at variance with a previous {\em ab initio} study based on the fully
anisotropic gap equations by Choi {\it et al.} \cite{choi3}, where doping was 
modeled by simply introducing excess electrons. For a moderate doping level of 
$x=0.2$ ($y=0.1$), they found a severe degradation of the $\pi$ gap while the 
$\sigma$ gap was more robust. This failure of a rigid-band-like type approach 
indicates that a more self-consistent site-dependent treatment of the doping is 
required for a proper description of the superconducting properties in doped 
MgB$_2$.

Two previous computational studies \cite{kortus3,umma} of the superconducting 
properties of MgB$_2$ alloys adopted a scaling scheme to describe the doping 
dependence. The Eliashberg functions for the undoped compound were scaled 
taking into account the doping dependence of $N(E_F)$ and of the $E_{2g}$ 
phonon frequency.
As such an approach does not discriminate between the doping sites, Kortus
{\it et al.} \cite{kortus3} argued that the differences observed for Al and C 
doping are due to a larger interband scattering for C than for Al doping.
The present study, however, demonstrates that the difference between Al and C 
doping appears naturally within the VCA approach, without the need to introduce 
another free parameter such as the interband scattering, as long as the 
influence of doping on the structure and on the lattice dynamics is properly 
taken into account.

\section{CONCLUSIONS}
We have performed a {\em first-principles} study of the electron-phonon 
coupling and superconducting properties for the Mg$_{1-x}$Al$_x$B$_2$ and 
MgB$_{2(1-y)}$C$_{2y}$ alloys as a function of $x$ and $y$, respectively, by 
combining the self-consistent virtual-crystal approximation and the two-band 
Eliashberg model.
For undoped MgB$_2$, the Eliashberg function possess a main peak at around 70 
meV related to the $E_{2g}$-phonon mode coming from the $\sigma\sigma$ 
contribution, and a sharper peak at 90 meV which originates largely from the 
$\pi\pi$ contribution, and is related to the $B_{1g}$-phonon mode. The total 
coupling constant $\lambda_{tot}=0.67$ agrees with the experimental value of 
$0.65$ as deduced from tunneling measurements.
For the alloys we found that $\alpha^2F(\omega)$ depends very sensitively on 
doping. It exhibits pronounced changes both in shape and in position of its 
main peaks, which renders any attempt to derive it from the spectrum of undoped 
MgB$_2$ via scaling procedures very unreliable. The calculated evolution of the 
Eliashberg functions compare well with recent electron tunneling spectroscopy 
measurements on Al-doped thin films.
The e-ph coupling parameter and its different contributions decrease as a 
function of doping for both alloys. Although both Al and C dopants donate 
electrons, the e-ph coupling exhibits a clear dependence on the doping site, 
which is also reflected in $\alpha^{2}F(\omega)$. 
With the Coulomb pseudopotential fixed for the undoped compound, we could 
reproduce the experimental doping dependence of $\Delta_{\sigma}$, 
$\Delta_{\pi}$, and $T_c$ for both alloys. The observed differences between Al 
and C doping, like the doping range of superconductivity, are naturally 
obtained in the present VCA approach, without the need to invoke other factors, 
as, e.g., interband scattering.
These results emphasize that a quantitative description of the superconducting 
properties of the two MgB$_2$ alloys require a proper treatment of the doping 
at least on the level of VCA, and suggest that interband scattering plays only 
a minor role.

\begin{acknowledgments}
This research was supported by the Consejo Nacional de Ciencia y 
Tecnolog{\'\i}a (CONACYT, M{\'e}xico) under Grant No. 43830-F and the 
Karlsruher Institut f\"{u}r Technologie (KIT), Germany. One of the authors 
(O.P.-S.) gratefully acknowledges CONACYT-M\'exico and the Deutscher 
Akademischer Austausch Dienst (DAAD).
\end{acknowledgments}

\end{document}